\documentclass[10pt,letterpaper]{article}
\usepackage[top=0.85in, left=1in, right=1in, bottom=1in, footskip=0.75in, marginparwidth=2in]{geometry}

\usepackage[utf8]{inputenc}
\usepackage{cite}

\usepackage{amsmath} 
\usepackage{nameref,hyperref}

\usepackage[right]{lineno}

\usepackage{microtype}
\DisableLigatures[f]{encoding = *, family = * }
\usepackage{changepage}
\usepackage[aboveskip=1pt,labelfont=bf,labelsep=period,singlelinecheck=off]{caption}

\makeatletter
\renewcommand{\@biblabel}[1]{\quad#1.}
\makeatother

\usepackage{lastpage,fancyhdr,graphicx,multirow,amssymb}
\usepackage{epstopdf}
\fancyheadoffset[L]{0in} 
\fancyfootoffset[L]{0in}

\usepackage{color}
\definecolor{Gray}{gray}{.25}
\usepackage{graphicx}
\usepackage{sidecap}

\usepackage{wrapfig}
\usepackage[pscoord]{eso-pic}
\usepackage[fulladjust]{marginnote}
\reversemarginpar

\begin{document}
\vspace*{0.35in}

\begin{flushleft}
{\Large
\textbf\newline{PatchProt: Hydrophobic patch prediction using protein foundation models}
}
\newline

Dea Gogishvili \textsuperscript{1,2,*},
Emmanuel Minois-Genin \textsuperscript{1},
Jan van Eck \textsuperscript{2},
and Sanne Abeln \textsuperscript{1,2}
\\
\bigskip
\bf{1} Bioinformatics, Computer Science Department, Vrije Universiteit Amsterdam, Amsterdam, Netherlands
\\
\bf{2} AI Technology for Life, Department of Computing and Information Sciences, Department of Biology, Utrecht University, Utrecht, Netherlands
\\
\bigskip
*Corresponding author. E-mail: d.gogishvili@vu.nl

\end{flushleft}

\section*{Abstract}
Hydrophobic patches on protein surfaces play important functional roles in protein-protein and protein-ligand interactions. Large hydrophobic surfaces are also involved in the progression of aggregation diseases. Predicting exposed hydrophobic patches from a protein sequence has been shown to be a difficult task. Fine-tuning foundation models allows for adapting a model to the specific nuances of a new task using a much smaller dataset. Additionally, multi-task deep learning offers a promising solution for addressing data gaps, simultaneously outperforming single-task methods. In this study, we harnessed a recently released leading large language model ESM-2. Efficient fine-tuning of ESM-2 was achieved by leveraging a recently developed parameter-efficient fine-tuning method. This approach enabled comprehensive training of model layers without excessive parameters and without the need to include a computationally expensive multiple sequence analysis. We explored several related tasks, at local (residue) and global (protein) levels, to improve the representation of the model. As a result, our fine-tuned ESM-2 model, PatchProt, cannot only predict hydrophobic patch areas but also outperforms existing methods at predicting primary tasks, including secondary structure and surface accessibility predictions. 
Importantly, our analysis shows that including related local tasks can improve predictions on more difficult global tasks. This research sets a new standard for sequence-based protein property prediction and highlights the remarkable potential of fine-tuning foundation models enriching the model representation by training over related tasks.\\
\\
\bigskip
\textbf{Availability and implementation:} \url{https://github.com/Deagogishvili/chapter-multi-task}
\\
\bigskip
\textbf{keywords:} Multi-task learning, Protein property prediction, Protein language model, ESM, LoRA.\\

%\linenumbers

\section*{Introduction}

Predicting the largest hydrophobic patch (LHP) area on the protein surface is a complex learning task \cite{van2022sticky}. Proteins typically hide hydrophobic residues within their core to avoid interaction with water, a phenomenon known as the hydrophobic effect \cite{Dill1985, Dill1990}. When such \emph{sticky} residues appear on the surface, they can play key roles in functional protein-protein, -ligand, or -membrane interactions \cite{Chothia1975, young1994role, malleshappa2014}, as well as induce amyloid fibril formation in the context of aggregation diseases \cite{Iadanza2018, Tuttle2016, Chiti2006}. Keeping these residues internal is thought to be a key strategy to avert protein aggregation \cite{Dobson2003, Abeln2008, abeln2011accounting}. Hydrophobic areas on the surface of the protein can influence experimental processes, such as gel formation, protein crystallisation \cite{Wright1999}, and separation techniques \cite{moruz2017peptide}. LHPs can be used to identify aggregation-prone regions \cite{sankar2018aggscore} which pose significant hurdles for the development of therapeutic proteins, such as monoclonal antibodies \cite{m2017good, sankar2018aggscore}. Importantly, predicting the exposure of hydrophobic residues on the protein surface is not a trivial problem. Traditional methods predict the majority of hydrophobic residues to be fully buried \cite{kyte1982, van2022sticky}. The continued evolution of the tools and methodologies is needed to deepen our understanding of protein hydrophobicity, especially in the context of neurodegenerative diseases.

The ability to predict structural and functional protein properties directly from a primary sequence is of paramount importance for unravelling its function in the absence of experimental structural information or predictions of low confidence. Various computational tools mostly focus on either local (per residue) or global (protein level) predictions, by taking a protein sequence as input and outputting a value or class per amino acid or protein chain \cite{hou2022ten}. Typical local tasks are the prediction of secondary structural elements, backbone geometry, post-translational modifications, and residues on protein-protein interfaces \cite{klausen2019netsurfp, capel2022multi}. Properties, such as cellular localisation, expression levels, and functional annotations are mostly predicted at the global level \cite{almagro2017deeploc, waury2024proteome}. Due to the lack of local annotations for training, many prediction methods focus on tasks at the global level. Nevertheless, these global prediction tasks form a class of hard prediction problems, including solubility, aggregation propensity, stability, turnover, and LHPs \cite{khurana2018deepsol, van2022sticky, housmans2023guide}. While local values can typically be summarised as global values, the reverse process is typically not possible. Multi-task deep learning architectures were previously shown useful to enrich a model representation, where there is a scarcity of annotated data for the task of interest \cite{capel2022multi}. Fine-tuning foundation models, which have been pre-trained on a vast amount of data, allows for effectively adapting a model to a new task even with limited datasets \cite{hoie2022netsurfp, perez2023aggbert}. In this work we combine both approaches to train a well-performing model to predict global hydrophobicity measures; specifically, we explore if it is possible to train a model on both global and local tasks, allowing the model to learn a shared representation.

Machine learning has long leveraged evolutionary profiles in multiple sequence alignments (MSA) \cite{rost1994phd}, to predict local or global protein features, including 3-dimensional (3D) structure \cite{camacho2009blast, klausen2019netsurfp, jumper2021highly}. Generating an MSA is typically a rate-limiting step as it involves an exhaustive search of homologs \cite{camacho2009blast, remmert2012hhblits, potter2018hmmer, mirdita2019mmseqs2}. Since the development of transformer-based models \cite{vaswani2017attention} large language models (LLMs) have revolutionised the field of natural language processing (NLP) \cite{chowdhary2020natural} and have been successfully applied to the analysis of protein sequences \cite{elnaggar2021prottrans}. The MSA can now be partially captured by a protein language model (PLM) leading to an order-of-magnitude acceleration of high-resolution structure prediction \cite{heinzinger2019modeling}. Evolutionary Scale Models (ESM) developed by Meta were trained on predicting masked residues in protein sequences and have recently presented promising results in protein folding prediction \cite{lin2023evolutionary}. NetSurfP-2.0 is a state-of-the-art tool for protein secondary structure, solvent accessibility, and disorder from its primary sequence \cite{hoie2022netsurfp}. The replacement of time-consuming MSAs with the ESM-1b PLM \cite{rives2021biological} significantly decreased the runtime without compromising prediction accuracy (version 3.0) \cite{klausen2019netsurfp}. 

This study builds upon protein foundation models, incorporating advancements in PLMs through the use of the recently published ESM-2. Parameter-efficient fine-tuning methods enabled us to comprehensively train our models, ultimately outperforming state-of-the-art methods for primary tasks. We have further extended existing datasets with local and global (L)HP annotations and not only improved the global LHP predictions but also obtained the first model that can predict (L)HPs on a residue level. Moreover, our model was trained on other biologically relevant tasks, demonstrating the possibility of foundation models and multi-task strategies to improve the accuracy of protein property predictions even with sparse datasets.

\section*{Methods}

\subsection*{Standard dataset}
To benchmark the performance of PatchProt for the primary prediction tasks, training and test datasets were obtained from previous work and used to develop NetSurfP-2.0 and -3.0 \cite{klausen2019netsurfp}.  
The curated training dataset contains 10,848 proteins retrieved from PDB with a sequence similarity $\le$ 25\%. Test datasets CASP12 (n=21), CB513 (n=513) and TS115 (n=115) are classic datasets for evaluating protein feature prediction models. All residues in each chain in the training dataset are annotated by an eight-state secondary structure (Q8), three-state secondary structure (Q3), relative solvent-accessible area (RSA), absolute solvent-accessible area (ASA), and $\phi$ and $\psi$ dihedral angles with the DSSP software. Residues present in the chain RefSeq sequence, but not in the solved structure, were defined as disordered (no atomic coordinates are available for these residues) \cite{kabsch1983dictionary, klausen2019netsurfp}.

\subsection*{Dataset expansion}
To investigate auxiliary tasks, we extended the datasets described above with more features, including (L)HP area, normalised RNA expression, and species, ultimately combining residue-based and global protein properties (Table \ref{tabsupp:datasets}). For LHP annotations, we utilised the tool MolPatch to calculate the area of hydrophobic patches based on the 3D structure of all protein chains \cite{van2022sticky}. To calculate hydrophobic patches, PDB structures in the existing datasets were retrieved. During the dataset expansion, amino acid sequences of PDB structures were checked with the amino acid sequences of the NetSurfP dataset and entries with a sequence match of more than 95\% amino acids were selected and annotated (in total, 10,594 chains). The LHP global indicates the size of the largest hydrophobic patch for the whole chain, while (L)HP local depicts a binary annotation per amino acid, whether or not a specific residue is in a (L) hydrophobic patch. Previously, we have shown that only lowly expressed human proteins (based on mRNA expression), are predicted to have large hydrophobic patches \cite{van2022sticky}. Here, we explore whether adding normalised expression values would aid LHP predictions. For normalised expression annotations, RNA consensus tissue gene data were obtained from the human protein atlas \cite{uhlen2015tissue} as described in the recent study \cite{van2022sticky}. To obtain a single expression value for every gene, the highest expression value was selected among all the tissues in which each gene was expressed. To obtain distinct groups, obtained values were grouped and the two lowest and the two highest deciles were selected for the prediction task (in total 618 chains). Additionally, we assigned labels to proteins based on the ten most common species including it as a prediction task. These labels encompass the ten most prevalent species, including: \emph{H. sapiens} (n=1638), \emph{E. coli} (n=615), \emph{S. cerevisiae} (n=393), \emph{M. musculus} (n=291), \emph{B. subtilis} (n=153), \emph{M. tuberculosis} (n=144), \emph{P. aeruginosa} (n=143), \emph{T. thermophilus} (n=135), \emph{A. thaliana} (n=133), and \emph{T. maritima} (n=122). Importantly, the added annotations do not completely cover our training and test datasets. To handle missing values we ignored the loss value of the missing annotations in the multi-task loss (see the multi-task loss section in Methods).

To benchmark global (protein level) predictions for the largest hydrophobic patches, we used the same test set of monomeric proteins as described previously \cite{van2022sticky}. We checked the overlap with the training dataset and removed all proteins with the matching PDB identifiers. The final test dataset for THSA, RHSA and LHP predictions consisted of 346 monomeric proteins. To assess model predictions, relative error threshold curves were calculated as described previously \cite{van2022sticky}.

\subsection*{Model}

\begin{figure*}
\centering
\includegraphics[width=16cm]{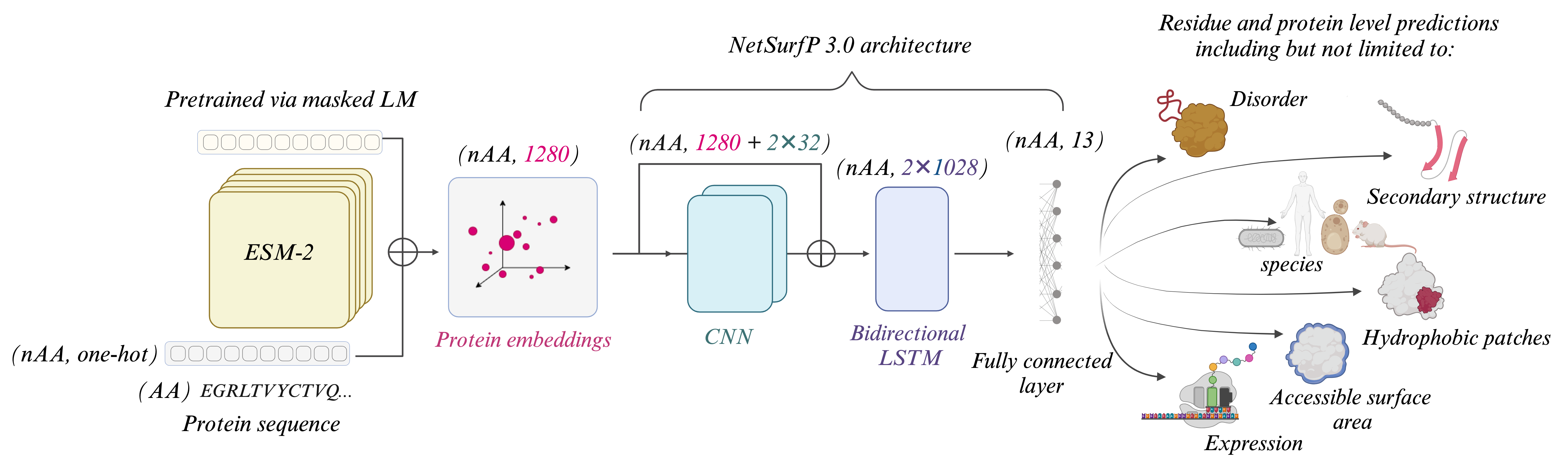}
\caption{\textbf{Model architecture.} The model takes protein sequence as input and predicts both global and local protein properties. The model consists of an embedding output from ESM-2 protein language model \cite{lin2023evolutionary} and the downstream architecture similar to NetSurfP-3.0 \cite{hoie2022netsurfp}. Additionally, a parameter-efficient fine-tuning strategy was implemented (Figure \ref{figsupp:LoRA}) \cite{hu2021LoRA, pfeiffer2021adapterfusion}. The decoding head consists of a residual block with two convolutional neural network (CNN) layers and a two-layer bidirectional long short-term memory (BiLSTM) network. The output is fed into a fully connected layer to provide predictions for all residues- and protein-level tasks.}\label{overview}
\end{figure*}

Our approach to building a deep learning model architecture was inspired by NetSurfP-3.0 \cite{hoie2022netsurfp}. In addition, we implemented an efficient fine-tuning strategy and explored a wide range of related global and local tasks. Figure \ref{overview} shows the overview of the model architecture. We utilised the embedding output from ESM-2 PLM \cite{lin2023evolutionary} and applied the downstream architecture to obtain predictions. Detailed information about the dimensions of the model, fine-tuning by Low-Rank Adaptation (LoRA), batch size correction, and our strategy of combining global and local tasks can be found in the Supplementary Information.

\subsection*{Multi-task loss}

Multi-task learning is a powerful approach in machine learning where a model is trained on multiple tasks simultaneously, leveraging commonalities and differences across tasks to obtain robust representations and improve generalisation \cite{liu2019endtoend, kendall2018multitask}. However, one of the key challenges in multi-task learning is effectively balancing the learning across tasks, as each task may have different levels of difficulty and importance. This necessitates the development of strategies to dynamically adjust the emphasis on each task during the training process.

The uncertainty-based loss, described by Liebel \emph{et. al.,} has shown promise in dynamically balancing the contribution of different tasks based on their levels of uncertainty by weighting each with a factor $\sigma_t$ \cite{liebel2018auxiliary}. To calculate individual losses, mean squared loss (RSA, $\phi$, $\psi$, TASA, THSA global LHP) and cross-entropy loss (Q8, Q3, disorder, local (L)HP, species, expression) are used. The multi-task loss function is then defined as:

\begin{align}
L_{multi}=\sum_{t\in \tau}\frac{L_t}{2\sigma_t^2}+\ln(1+\sigma_t^2),\label{eq1}
\end{align}

where $\tau$ represents the set of tasks, $L_t$ is the loss for task $t$, and $\sigma_t$ is the uncertainty associated with each task in a multi-task learning setting $t$ (see Supplementary Information). To prevent negative loss values $\ln(1+\sigma_t^2)$ is administered in the approach and the last term acts as a regulariser. The summation of these components across all tasks $\tau$, where $\tau$ includes only the tasks with non-null values in a given batch, forms the final loss function. This formulation allows the model to prioritise tasks based on their current level of uncertainty, potentially leading to more effective and efficient learning. Notably, when using LoRA for efficient fine-tuning, the multi-task loss is applied to the whole architecture (Figure \ref{overview}) (in this case, LoRA introduces low-rank matrices that are trainable and are added to the pre-existing weights of the ESM-2 model). If there is no fine-tuning chosen, then the multi-task loss is applied to the CNN-LSTM part of the model following the generation of embeddings (as depicted in Figure \ref{overview}). Detailed information about scaling losses according to uncertainty can be found in the supplementary information.

\begin{table*}[th!]
\begin{tabular}{@{\extracolsep\fill}lllllllllll@{\extracolsep\fill}}
\hline
\multirow{2}{*}{Test dataset} & \multirow{2}{*}{Model} & RSA $\uparrow$  & ASA $\uparrow$  & Q8  $\uparrow$ & Q3  $\uparrow$  & Dis $\uparrow$ & Dis $\downarrow$ & Phi $\downarrow$ & Psi $\downarrow$ \\
 & & (PCC) & (PCC) & (ACC) & (ACC) & (MCC) & (FNR)  & (MAE) & (MAE)\\
\hline

\multirow{4}{*}{CASP12}
& NetSurfP-2.0 &  0.728 & 0.739 & 0.699 & 0.810 & 0.653 &  \textbf{0.015} & 20.90 & 32.80\\
& NetSurfP-3.0 & 0.707 & 0.722 & 0.669 & 0.791 & 0.621 & 0.024 & 21.25 & 33.92 \\
& PatchProt* & \textbf{0.740} & \textbf{0.748} & \textbf{0.695} & \textbf{0.817} & \textbf{0.658} & 0.026 & \textbf{20.20} & \textbf{30.95} \\
& PatchProt** & 0.724 & 0.741 & 0.685 & 0.795 & 0.592 & 0.032 &  20.42 & 32.42 \\
\hline

\multirow{4}{*}{CB513}
& NetSurfP-2.0 & 0.791 & 0.804 & 0.713 & 0.845 & - & - & 20.35 & 29.04\\
& NetSurfP-3.0 & 0.793 & 0.810 & 0.711 & 0.846 & - & - & 20.22 & 29.25\\
& PatchProt* & 0.811 & 0.823 & 0.724 & 0.860 & - & - & 19.47 & 26.73 \\
& PatchProt** & \textbf{0.816} & \textbf{0.828} & \textbf{0.738} &  \textbf{0.868} & - & - & \textbf{18.83} & \textbf{25.72} \\
\hline

\multirow{4}{*}{TS115}
& NetSurfP-2.0 & 0.771 & 0.793 & 0.740 & 0.849 & 0.624 & \textbf{0.013} & 17.40 & 26.80 \\
& NetSurfP-3.0 & 0.776 & 0.799 & 0.749 & 0.856 &  0.662 & 0.015 & 17.16 & 25.80 \\
& PatchProt* & 0.796 & 0.812 & 0.757 & 0.867  & \textbf{0.667} & 0.016 & 16.67 & 23.75 \\
& PatchProt** &  \textbf{0.799} & \textbf{0.817} & \textbf{0.765} & \textbf{0.871} & 0.649 & 0.016 & \textbf{16.24} & \textbf{23.67} \\
\hline

\end{tabular}
\caption{\textbf{Model performance when applying ESM-2 embeddings to predict protein local structure.} Comparison of NetSurfP-2.0, NetSurfP-3.0, and our model - PatchProt on the CB513, TS115 and CASP12 datasets. * PatchProt (only secondary structure element, without LoRA). ** PatchProt (all tasks, with LoRA). Performance values for the NetsurfP models are reported as stated in the latest publication \cite{hoie2022netsurfp}. Dis - disorder.  Each column reports an output variable with the same corresponding metrics reported in the previous study \cite{hoie2022netsurfp} for benchmarking purposes: Pearson correlation coefficient (PCC), accuracy (ACC), Matthews correlation coefficient (MCC), false negative rate (FNR) and mean absolute error (MAE). Up- and down-facing arrows indicate metrics for which an improvement represents larger or lower values. For each dataset and prediction task, the values corresponding to the best performance are shown in bold. A complete table for the comparison of all the models with and without fine-tuning can be found in the supplementary information (Table \ref{tabsupp:basic_tasks_SI}).
\label{basic_tasks}}%
\end{table*}

\section*{Results}\label{sec4}

In this study, we first aimed to explore the potential of foundation models and efficient fine-tuning strategies to improve primary protein property prediction tasks. Second, using multi-task learning, we set to obtain a well-performing (L)HP predictor on both residue and protein levels. Finally, we expanded the model and added auxiliary tasks with scarce annotations to ascertain the possibility of using limited datasets to take advantage of shared representations. 

\subsection*{Improved secondary structure predictions}

To improve the model architecture, we first explored a recent release of the ESM-2 language model by META. This allowed us to outperform previous well-established sequence-based models NetSurfP versions 2.0 and 3.0 on typical tasks, including RSA, ASA, and geometry. Tables \ref{basic_tasks} and \ref{tabsupp:basic_tasks_SI} show the comparison of different models on typical secondary structure component predictions across different test datasets. NetSurfP-3.0 is a model that consists of the embedding model ESM-1b with the NetSurfP-3.0 head (ResNet encoder and bidirectional LSTM) at its end. The architecture of our model - PatchProt combines the ESM-2 and the NetSurfP-3.0 head. Additionally, our model includes an efficient fine-tuning strategy and a wide range of related global and local tasks. ESM-2 alone achieves similar performance as NetSurfP 3.0 and outperforms it in almost all tasks (Table \ref{tabsupp:basic_tasks_SI}). PatchProt with LoRA and auxiliary tasks leads to improved predictions in primary tasks (Table \ref{basic_tasks}). 

\subsection*{Improved large hydrophobic patch predictions}

\begin{figure*}[b!]
\centering
\includegraphics[width=16cm]{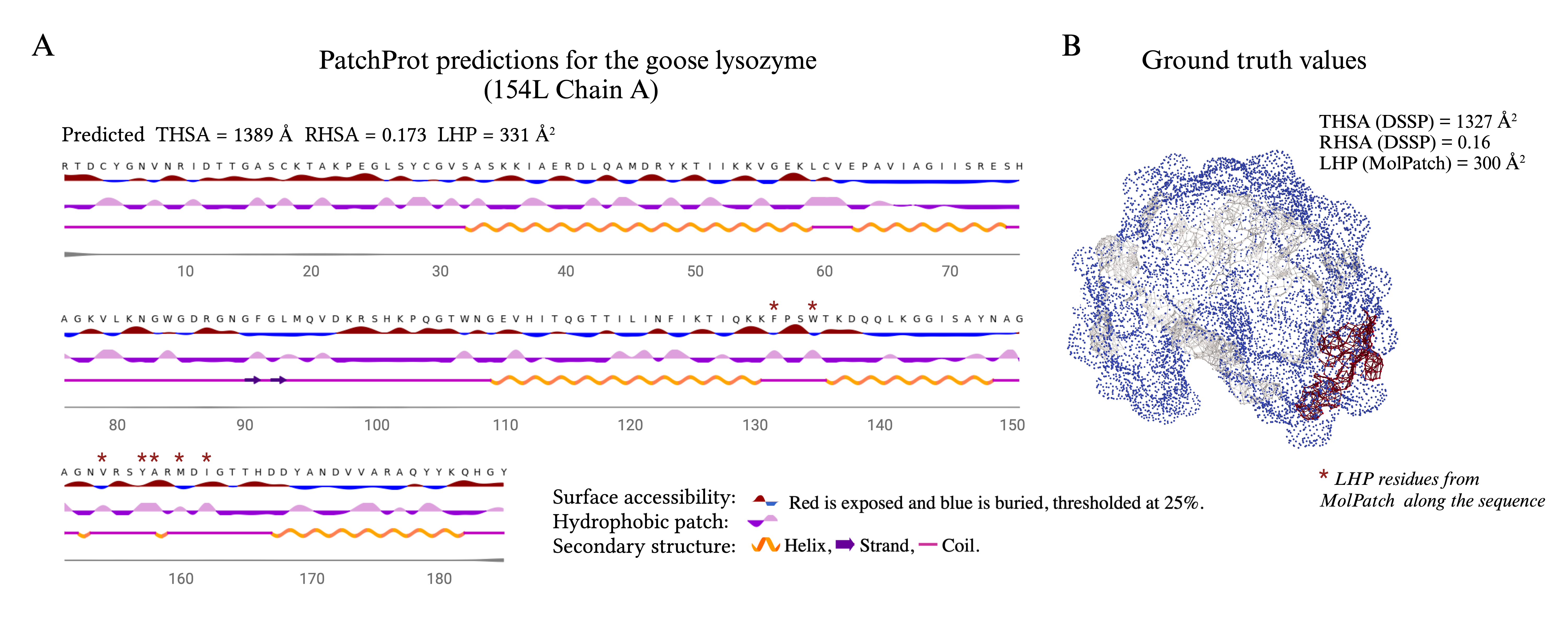}
\caption{\textbf{Assessment of hydrophobic patch (HP) predictions.} (\textbf{A}) 154L chain A - Case example from the test set of CB513. A visualisation for PatchProt predictions in a manner of NetSurfP-3.0. (\textbf{B}) Ground truth labels for the same protein structure were calculated from DSSP (for total hydrophobic surface area (THSA) and relative hydrophobic surface area (RHSA) and MolPatch (for the largest HP).}
\label{case_example}
\end{figure*}

It has been previously shown that predicting the global largest hydrophobic patch area for a protein is not a trivial task \cite{van2022sticky}. By training shallow learning algorithms on basic protein characteristics, such as sequence length, and number of hydrophobic and hydrophilic residues, the performance of the model was relatively low ($R^{2}$ = 0.12). When incorporating TASA and THSA values predicted by NetSurfP-2.0, the performance improved ($R^{2}$ = 0.43). Here we not only achieved a significantly higher performance on the global level, but we also added residue-based predictions that to the best of our knowledge, have not been attempted before (Table \ref{tab_added_tasks}). Moreover, it is possible to visualise the hydrophobic patches at a residue level in a similar manner as implemented for NetsurfP-3.0 (Figure \ref{case_example}A) \cite{hoie2022netsurfp}. A case example was randomly selected from the test set of CB513 and the predictions by PatchProt are comparable to the ground truth LHP area calculated by MolPatch (Figure \ref{case_example}).

To assess the model performance on added tasks, we compare three models: \textit{(i)} The \emph{(L)HP} only model, which is trained on predicting hydrophobic patches (both global and local) without any additional features. \textit{(ii)} The SSE + (L)HP model and \textit{(iii)} the final model that includes all the implemented tasks to explore whether adding less relevant global tasks would improve or worsen the performance. The \emph{(L)HP only} model performed significantly worse than every other model. When we added (L)HP tasks to the primary secondary structure properties, the (L)HP predictions improved suggesting the benefits of a multi-task learning strategy. Adding normalised expression values, and species improved the global LHP predictions, however, we did not observe a significant added benefit of global tasks in the residue-level performance measures.

\begin{table*}[t!]
\begin{tabular}{@{\extracolsep\fill}lllllllll@{\extracolsep\fill}}
\hline
\multirow{2}{*}{Test dataset} & PatchProt & LHP g $\downarrow$ & HP l $\uparrow$ & HP l $\downarrow$ & LHP l $\uparrow$ & LHP l $\downarrow$ & SP $\uparrow$ & NX $\uparrow$ \\
 & prediction tasks & (MAE) & (MCC) & (FNR) & (MCC) & (FNR) & (ACC) & (ACC) \\
\hline

\multirow{3}{*}{CASP12}
& (L)HP only & 588.9  & 0.854 & 0.070 & 0.405 & 0.682 & - & - \\
& SSE, (L)HP & 497.0 & 0.855 & \textbf{0.047} & 0.397 & \textbf{0.619} & - & - \\
& SSE, (L)HP, SP, NX & \textbf{449.8} & \textbf{0.858} & 0.053 & \textbf{0.461} & 0.706 & 1 & 1 \\

\hline

\multirow{3}{*}{CB513}
& (L)HP only & 434.1 & 0.861 & 0.072 & 0.369 & 0.681 & - & - \\
& SSE, (L)HP & 418.7 & \textbf{0.865} & \textbf{0.048} & \textbf{0.392} & \textbf{0.630} & - & - \\
& SSE, (L)HP, SP, NX & \textbf{416.7} & 0.864 & 0.059 & 0.335 & 0.729 & 0.683 & 0.269 \\

\hline

\multirow{3}{*}{TS115}
& (L)HP only & \textbf{483.8} & 0.866 & 0.063 & 0.375 & 0.685 & - & -  \\
& SSE, (L)HP & 503.8 & 0.869 & \textbf{0.045} & \textbf{0.419} & \textbf{0.603} & - & - \\
& SSE, (L)HP, SP, NX & 517.8 & \textbf{0.870} & 0.054 & 0.342 & 0.726 & 0.745 & 0.857 \\

\hline

\end{tabular}
\caption{Model performance for additional local (l) and global (g) tasks. Performance of our multi-task model on the CB513, TS115 and CASP12 datasets compared with the multi-task and single-task models. Each column reports an output variable with the corresponding metrics: Accuracy (ACC), Matthews correlation coefficient (MCC), false negative rate (FNR), and global mean absolute error (MAE). SSE - Secondary structure element, primary tasks (Table \ref{basic_tasks}), LHP - largest hydrophobic patch, NX - normalised expression, SP - species. Up- and down-facing arrows indicate metrics for which an improvement represents larger or lower values. For each dataset and prediction task, the values corresponding to the best performance are shown in bold.
\label{tab_added_tasks}}
\end{table*}

To assess the performance of PatchProt on global predictions for LHP values, we benchmarked our predictions against other methods reported previously (Figure \ref{figsupp:threshold_curve}) \cite{van2022sticky}. For difficult regression tasks, $R^{2}$ or the mean absolute error (MAE) values are heavily influenced by outliers and do not generally produce results that are easy to interpret. In addition to the $R^{2}$ and MAE metrics, we evaluated the performance of the prediction model by examining the relative error threshold curve given a certain threshold, inspired by the GDT-TS score \cite{zemla2001processing, van2022sticky}. PatchProt achieved better performance for global LHP values than all the other models developed before (Figure \ref{figsupp:threshold_curve}).

\section*{Discussion and Conclusions}

In this article, we present an approach to fine-tune LLM for multi-task protein property prediction. Our method outperformed currently published best-performing models in well-established secondary structure component prediction tasks without a time-consuming MSA step (Table \ref{basic_tasks}).
An exhaustive search of homologs is a rate-limiting step for MSA-based methods and it can now be partially captured by PLMs leading to a substantial acceleration of predictions \cite{camacho2009blast, remmert2012hhblits, potter2018hmmer, mirdita2019mmseqs2, heinzinger2019modeling}. We believe that our improvement is possible by the pre-trained ESM2 model used to encode the language of proteins, a recently published PLM by Meta which outperforms every other PLM on a wide variety of tasks. By solely changing the PLM from ESM-1b to ESM-2, we observe an increase in overall performance (Table \ref{tabsupp:basic_tasks_SI}). Even without the downstream architecture of NetSurfP-3.0 (convolutional encoder and bi-directional LSTM), PatchProt outperforms both NetSurfP-2.0 and -3.0 in most tasks (Table \ref{tabsupp:basic_tasks_SI}).

In addition to the local residue-based tasks, our model - PatchProt can predict global tasks. With this approach, we can combine relevant local and global tasks and significantly improve global predictions that are challenging otherwise \cite{capel2022multi}. We have demonstrated that PatchProt can leverage shared representations from other related tasks and achieve strong performance in predicting hydrophobic patches (Table \ref{tab_added_tasks}). Importantly, global LHP predictions were previously demonstrated to be challenging \cite{van2022sticky} and local LHP predictions, to the best of our knowledge, have never been attempted. Additionally, we have shown that adding less relevant tasks (expression and species) does not harm the model performance in primary prediction tasks and can even improve the performance in certain cases (Table \ref{tabsupp:basic_tasks_SI}). Additionally, we obtain better predictions on added tasks through multi-tasking compared to single-task models demonstrated by hydrophobic patches. Often, biologically relevant predictions suffer from low-quality or less standardised datasets. Here we have demonstrated, that data scarcity could be tackled by combining existing datasets with limited annotations to benefit from commonalities among the prediction tasks.

AlphaFold has greatly advanced our potential to utilise deep-learning methods for predicting protein structures from sequences \cite{jumper2021highly} allowing open access to over 200 million protein structure predictions \cite{varadi2022alphafold}. Nevertheless, structure-based methods might not be as accurate on predicted models as on respective structures determined by X-ray crystallography. One of the challenges for using predicted 3D structures by AlphaFold is linked to disordered regions. Having larger surface accessibility in coiled regions or non-globular proteins, in general, can lead to overestimated LHP area calculations. Moreover, unlike AlphaFold, here we focus on predicting LHPs without the need to have an MSA.

Predicting protein properties directly from amino acid sequences is a valuable way to quickly and accurately annotate proteins. While protein foundation models offer an outstanding opportunity to improve predictions on various challenging tasks, the memory requirements of LLMs can be a significant limitation for their use on resource-constrained devices. Ongoing research in quantised LLMs is expected to further enhance their applicability in protein science applications. Quantisation schemes such as fixed-point representation, weight sharing, and Huffman coding \cite{han2015deep, goyal2021fixed} can be used to reduce the memory footprint of LLMs by representing the model parameters using fewer bits \cite{li2023loftq}. 

To summarise, with our model architecture that combines an advanced fine-tuning strategy with related task prediction, we not only demonstrate the possibility to outperform state-of-the-art tools in established secondary structure element predictions but also to add prediction tasks that are challenged with data scarcity or intrinsic difficulty. Specifically, our analysis shows that including related residue-level tasks can improve performance on more difficult global tasks, such as LHP areas. Our approach can be applied to other complex global properties, such as turnover, aggregation-propensity, and solubility prediction tasks. Continued research in LLMs will further enhance the effectiveness and applicability of fine-tuning and leveraging powerful representations of protein sequences for various relevant tasks.

\section*{Acknowledgments}
We would like to thank Wilson Silva for the critical review of the manuscript.
The authors acknowledge BAZIS HPC cluster computing facilities at the Vrije Universiteit Amsterdam.

\section*{Funding}
Research of DG and SA is supported by the European Commission (Marie Curie International Training Network, grant agreement No 860197 (MIRIADE). The research of SA is supported by Health-Holland.

\section*{Competing interests}
Outside the submitted work: SA reports a patent pending; SA is in a consortium agreement with Cergentis BV as part of the TargetSV project; SA is in a consortium agreement with Olink and Quanterix as part of the NORMAL project. The rest of the authors have no competing interests to declare.

\section*{Author contributions statement}

Conceptualisation: DG, EMG, SA;
Data collection: DG, EMG, JvE;
Data curation: DG, EMG, JvE;
Methodology: DG, EMG, JvE;
ML architecture and fine-tuning strategy: EMG;
Formal analysis: DG, EMG, JvE;
Funding acquisition: SA;
Supervision: DG, SA
Validation: DG, JvE;
Visualisation: DG, JvE, EMG;
Writing - original draft preparation: DG, EMG, JvE, SA;
Writing - review \& editing: DG, EMG, JvE, SA.

\section*{Data availability}
Data and code implemented in this study are available at: \url{https://github.com/Deagogishvili/chapter-multi-task}.

\nolinenumbers

\bibliography{main}

\begin{thebibliography}{10}

\bibitem{van2022sticky}
J.~H.~M. van Gils, D.~Gogishvili, J.~van Eck, R.~Bouwmeester, E.~van Dijk, and S.~Abeln, ``How sticky are our proteins? quantifying hydrophobicity of the human proteome,'' {\em Bioinformatics advances}, vol.~2, no.~1, p.~vbac002, 2022.

\bibitem{Dill1985}
K.~A. Dill, ``Theory for the folding and stability of globular proteins,'' {\em Biochemistry}, vol.~24, pp.~1501--1509, 1985.

\bibitem{Dill1990}
K.~A. Dill, ``Dominant forces in protein folding,'' {\em Biochemistry}, vol.~29, pp.~7133--7155, 1990.

\bibitem{Chothia1975}
C.~Chothia and J.~Janin, ``Principles of protein–protein recognition,'' {\em Nature}, vol.~256, pp.~705--708, 1975.

\bibitem{young1994role}
L.~Young, R.~Jernigan, and D.~Covell, ``A role for surface hydrophobicity in protein-protein recognition,'' {\em Protein Science}, vol.~3, no.~5, pp.~717--729, 1994.

\bibitem{malleshappa2014}
S.~M. Gowder, J.~Chatterjee, T.~Chaudhuri, and K.~Paul, ``Prediction and analysis of surface hydrophobic residues in tertiary structure of proteins,'' {\em The Scientific World Journal}, vol.~2014, 2014.

\bibitem{Iadanza2018}
M.~G. Iadanza, R.~Silvers, J.~Boardman, H.~I. Smith, T.~K. Karamanos, G.~T. Debelouchina, Y.~Su, R.~G. Griffin, N.~A. Ranson, and S.~E. Radford, ``{The structure of a $\beta$2-microglobulin fibril suggests a molecular basis for its amyloid polymorphism},'' {\em Nature Communications}, vol.~9, p.~4517, Dec. 2018.

\bibitem{Tuttle2016}
M.~D. Tuttle, G.~Comellas, A.~J. Nieuwkoop, D.~J. Covell, D.~A. Berthold, K.~D. Kloepper, J.~M. Courtney, J.~K. Kim, A.~M. Barclay, A.~Kendall, W.~Wan, G.~Stubbs, C.~D. Schwieters, V.~M.~Y. Lee, J.~M. George, and C.~M. Rienstra, ``{Solid-state NMR structure of a pathogenic fibril of full-length human [alpha]-synuclein},'' {\em Nat Struct Mol Biol}, vol.~23, pp.~409--415, May 2016.

\bibitem{Chiti2006}
F.~Chiti and C.~M. Dobson, ``Protein misfolding, functional amyloid, and human disease,'' {\em Annu. Rev. Biochem.}, vol.~75, pp.~333--366, 2006.

\bibitem{Dobson2003}
C.~M. Dobson, ``Protein folding and disease: a view from the first horizon symposium,'' {\em Nature Reviews Drug Discovery}, vol.~2, pp.~154--160, 2003.

\bibitem{Abeln2008}
S.~Abeln and D.~Frenkel, ``{Disordered Flanks Prevent Peptide Aggregation},'' {\em PLoS Comput Biol}, vol.~4, no.~12, p.~e1000241, 2008.

\bibitem{abeln2011accounting}
S.~Abeln and D.~Frenkel, ``Accounting for protein-solvent contacts facilitates design of nonaggregating lattice proteins,'' {\em Biophysical journal}, vol.~100, no.~3, pp.~693--700, 2011.

\bibitem{Wright1999}
P.~E. Wright and H.~J. Dyson, ``Intrinsically unstructured proteins: re-assessing the protein structure-function paradigm,'' {\em Journal of molecular biology}, vol.~293, pp.~321--331, 1999.

\bibitem{moruz2017peptide}
L.~Moruz and L.~K{\"a}ll, ``Peptide retention time prediction,'' {\em Mass spectrometry reviews}, vol.~36, no.~5, pp.~615--623, 2017.

\bibitem{sankar2018aggscore}
K.~Sankar, S.~R. Krystek~Jr, S.~M. Carl, T.~Day, and J.~K. Maier, ``Aggscore: Prediction of aggregation-prone regions in proteins based on the distribution of surface patches,'' {\em Proteins: Structure, Function, and Bioinformatics}, vol.~86, no.~11, pp.~1147--1156, 2018.

\bibitem{m2017good}
J.~M~Redington, L.~Breydo, and V.~N~Uversky, ``When good goes awry: the aggregation of protein therapeutics,'' {\em Protein and Peptide Letters}, vol.~24, no.~4, pp.~340--347, 2017.

\bibitem{kyte1982}
J.~Kyte and R.~F. Doolittle, ``A simple method for displaying the hydropathic character of a protein,'' {\em Journal of Molecular Biology}, vol.~157, no.~1, pp.~105 -- 132, 1982.

\bibitem{hou2022ten}
Q.~Hou, K.~Waury, D.~Gogishvili, and K.~A. Feenstra, ``Ten quick tips for sequence-based prediction of protein properties using machine learning,'' {\em PLOS Computational Biology}, vol.~18, no.~12, p.~e1010669, 2022.

\bibitem{klausen2019netsurfp}
M.~S. Klausen, M.~C. Jespersen, H.~Nielsen, K.~K. Jensen, V.~I. Jurtz, C.~K. Soenderby, M.~O.~A. Sommer, O.~Winther, M.~Nielsen, B.~Petersen, {\em et~al.}, ``Netsurfp-2.0: Improved prediction of protein structural features by integrated deep learning,'' {\em Proteins: Structure, Function, and Bioinformatics}, vol.~87, no.~6, pp.~520--527, 2019.

\bibitem{capel2022multi}
H.~Capel, K.~A. Feenstra, and S.~Abeln, ``Multi-task learning to leverage partially annotated data for ppi interface prediction,'' {\em Scientific Reports}, vol.~12, no.~1, p.~10487, 2022.

\bibitem{almagro2017deeploc}
J.~J. Almagro~Armenteros, C.~K. S{\o}nderby, S.~K. S{\o}nderby, H.~Nielsen, and O.~Winther, ``Deeploc: prediction of protein subcellular localization using deep learning,'' {\em Bioinformatics}, vol.~33, no.~21, pp.~3387--3395, 2017.

\bibitem{waury2024proteome}
K.~Waury, D.~Gogishvili, R.~Nieuwland, M.~Chatterjee, C.~E. Teunissen, and S.~Abeln, ``Proteome encoded determinants of protein sorting into extracellular vesicles,'' {\em Journal of Extracellular Biology}, vol.~3, no.~1, p.~e120, 2024.

\bibitem{khurana2018deepsol}
S.~Khurana, R.~Rawi, K.~Kunji, G.-Y. Chuang, H.~Bensmail, and R.~Mall, ``Deepsol: a deep learning framework for sequence-based protein solubility prediction,'' {\em Bioinformatics}, vol.~34, no.~15, pp.~2605--2613, 2018.

\bibitem{housmans2023guide}
J.~A. Housmans, G.~Wu, J.~Schymkowitz, and F.~Rousseau, ``A guide to studying protein aggregation,'' {\em The FEBS Journal}, vol.~290, no.~3, pp.~554--583, 2023.

\bibitem{hoie2022netsurfp}
M.~H. H{\o}ie, E.~N. Kiehl, B.~Petersen, M.~Nielsen, O.~Winther, H.~Nielsen, J.~Hallgren, and P.~Marcatili, ``Netsurfp-3.0: accurate and fast prediction of protein structural features by protein language models and deep learning,'' {\em Nucleic acids research}, vol.~50, no.~W1, pp.~W510--W515, 2022.

\bibitem{perez2023aggbert}
R.~Perez, X.~Li, S.~Giannakoulias, and E.~J. Petersson, ``Aggbert: Best in class prediction of hexapeptide amyloidogenesis with a semi-supervised protbert model,'' {\em Journal of Chemical Information and Modeling}, 2023.

\bibitem{rost1994phd}
B.~Rost, C.~Sander, and R.~Schneider, ``Phd-an automatic mail server for protein secondary structure prediction,'' {\em Bioinformatics}, vol.~10, no.~1, pp.~53--60, 1994.

\bibitem{camacho2009blast}
C.~Camacho, G.~Coulouris, V.~Avagyan, N.~Ma, J.~Papadopoulos, K.~Bealer, and T.~L. Madden, ``Blast+: architecture and applications,'' {\em BMC bioinformatics}, vol.~10, pp.~1--9, 2009.

\bibitem{jumper2021highly}
J.~Jumper, R.~Evans, A.~Pritzel, T.~Green, M.~Figurnov, O.~Ronneberger, K.~Tunyasuvunakool, R.~Bates, A.~{\v{Z}}{\'\i}dek, A.~Potapenko, {\em et~al.}, ``Highly accurate protein structure prediction with alphafold,'' {\em Nature}, vol.~596, no.~7873, pp.~583--589, 2021.

\bibitem{remmert2012hhblits}
M.~Remmert, A.~Biegert, A.~Hauser, and J.~S{\"o}ding, ``Hhblits: lightning-fast iterative protein sequence searching by hmm-hmm alignment,'' {\em Nature methods}, vol.~9, no.~2, pp.~173--175, 2012.

\bibitem{potter2018hmmer}
S.~C. Potter, A.~Luciani, S.~R. Eddy, Y.~Park, R.~Lopez, and R.~D. Finn, ``Hmmer web server: 2018 update,'' {\em Nucleic acids research}, vol.~46, no.~W1, pp.~W200--W204, 2018.

\bibitem{mirdita2019mmseqs2}
M.~Mirdita, M.~Steinegger, and J.~S{\"o}ding, ``Mmseqs2 desktop and local web server app for fast, interactive sequence searches,'' {\em Bioinformatics}, vol.~35, no.~16, pp.~2856--2858, 2019.

\bibitem{vaswani2017attention}
A.~Vaswani, N.~Shazeer, N.~Parmar, J.~Uszkoreit, L.~Jones, A.~N. Gomez, L.~Kaiser, and I.~Polosukhin, ``Attention is all you need,'' 2017.

\bibitem{chowdhary2020natural}
K.~Chowdhary and K.~Chowdhary, ``Natural language processing,'' {\em Fundamentals of artificial intelligence}, pp.~603--649, 2020.

\bibitem{elnaggar2021prottrans}
A.~Elnaggar, M.~Heinzinger, C.~Dallago, G.~Rehawi, Y.~Wang, L.~Jones, T.~Gibbs, T.~Feher, C.~Angerer, M.~Steinegger, {\em et~al.}, ``Prottrans: Toward understanding the language of life through self-supervised learning,'' {\em IEEE transactions on pattern analysis and machine intelligence}, vol.~44, no.~10, pp.~7112--7127, 2021.

\bibitem{heinzinger2019modeling}
M.~Heinzinger, A.~Elnaggar, Y.~Wang, C.~Dallago, D.~Nechaev, F.~Matthes, and B.~Rost, ``Modeling aspects of the language of life through transfer-learning protein sequences,'' {\em BMC bioinformatics}, vol.~20, pp.~1--17, 2019.

\bibitem{lin2023evolutionary}
Z.~Lin, H.~Akin, R.~Rao, B.~Hie, Z.~Zhu, W.~Lu, N.~Smetanin, R.~Verkuil, O.~Kabeli, Y.~Shmueli, {\em et~al.}, ``Evolutionary-scale prediction of atomic-level protein structure with a language model,'' {\em Science}, vol.~379, no.~6637, pp.~1123--1130, 2023.

\bibitem{rives2021biological}
A.~Rives, J.~Meier, T.~Sercu, S.~Goyal, Z.~Lin, J.~Liu, D.~Guo, M.~Ott, C.~L. Zitnick, J.~Ma, {\em et~al.}, ``Biological structure and function emerge from scaling unsupervised learning to 250 million protein sequences,'' {\em Proceedings of the National Academy of Sciences}, vol.~118, no.~15, p.~e2016239118, 2021.

\bibitem{kabsch1983dictionary}
W.~Kabsch and C.~Sander, ``Dictionary of protein secondary structure: pattern recognition of hydrogen-bonded and geometrical features,'' {\em Biopolymers: Original Research on Biomolecules}, vol.~22, no.~12, pp.~2577--2637, 1983.

\bibitem{uhlen2015tissue}
M.~Uhl{\'e}n, L.~Fagerberg, B.~M. Hallstr{\"o}m, C.~Lindskog, P.~Oksvold, A.~Mardinoglu, {\AA}.~Sivertsson, C.~Kampf, E.~Sj{\"o}stedt, A.~Asplund, {\em et~al.}, ``Tissue-based map of the human proteome,'' {\em Science}, vol.~347, no.~6220, p.~1260419, 2015.

\bibitem{hu2021LoRA}
E.~J. Hu, Y.~Shen, P.~Wallis, Z.~Allen-Zhu, Y.~Li, S.~Wang, L.~Wang, and W.~Chen, ``Lora: Low-rank adaptation of large language models,'' 2021.

\bibitem{pfeiffer2021adapterfusion}
J.~Pfeiffer, A.~Kamath, A.~Rücklé, K.~Cho, and I.~Gurevych, ``Adapterfusion: Non-destructive task composition for transfer learning,'' 2021.

\bibitem{liu2019endtoend}
S.~Liu, E.~Johns, and A.~J. Davison, ``End-to-end multi-task learning with attention,'' 2019.

\bibitem{kendall2018multitask}
A.~Kendall, Y.~Gal, and R.~Cipolla, ``Multi-task learning using uncertainty to weigh losses for scene geometry and semantics,'' 2018.

\bibitem{liebel2018auxiliary}
L.~Liebel and M.~Körner, ``Auxiliary tasks in multi-task learning,'' 2018.

\bibitem{zemla2001processing}
A.~Zemla, {\v{C}}.~Venclovas, J.~Moult, and K.~Fidelis, ``Processing and evaluation of predictions in casp4,'' 2001.

\bibitem{varadi2022alphafold}
M.~Varadi, S.~Anyango, M.~Deshpande, S.~Nair, C.~Natassia, G.~Yordanova, D.~Yuan, O.~Stroe, G.~Wood, A.~Laydon, {\em et~al.}, ``Alphafold protein structure database: massively expanding the structural coverage of protein-sequence space with high-accuracy models,'' {\em Nucleic acids research}, vol.~50, no.~D1, pp.~D439--D444, 2022.

\bibitem{han2015deep}
S.~Han, H.~Mao, and W.~J. Dally, ``Deep compression: Compressing deep neural networks with pruning, trained quantization and huffman coding,'' {\em arXiv preprint arXiv:1510.00149}, 2015.

\bibitem{goyal2021fixed}
R.~Goyal, J.~Vanschoren, V.~Van~Acht, and S.~Nijssen, ``Fixed-point quantization of convolutional neural networks for quantized inference on embedded platforms,'' {\em arXiv preprint arXiv:2102.02147}, 2021.

\bibitem{li2023loftq}
Y.~Li, Y.~Yu, C.~Liang, P.~He, N.~Karampatziakis, W.~Chen, and T.~Zhao, ``Loftq: Lora-fine-tuning-aware quantization for large language models,'' {\em arXiv preprint arXiv:2310.08659}, 2023.

\bibitem{kuhn2008building}
M.~Kuhn, ``Building predictive models in r using the caret package,'' {\em Journal of statistical software}, vol.~28, pp.~1--26, 2008.

\bibitem{chen2016xgboost}
T.~Chen and C.~Guestrin, ``Xgboost: A scalable tree boosting system,'' in {\em Proceedings of the 22nd acm sigkdd international conference on knowledge discovery and data mining}, pp.~785--794, 2016.

\end{thebibliography}

\bibliographystyle{ieeetr}

\clearpage
\renewcommand{\thefigure}{S\arabic{figure}}
\renewcommand{\theequation}{S\arabic{equation}}
\renewcommand{\thetable}{S\arabic{table}}

\setcounter{figure}{0}
\setcounter{equation}{0}
\setcounter{table}{0}

\onecolumn

\section*{Supplementary information}

\subsection*{Input and output dimensions}\label{dimensions}

To predict properties for a given protein, our input shape is defined as $(nAA, OH)$, where $nAA$ represents the number of amino acids in the protein and $OH$ signifies the one-hot-encoded amino acid at the respective index ($OH = 20$). For a given batch size, the input would be $(P, nAA, OH)$, where $P$ denotes the number of proteins in a single batch. PatchProt, utilising ESM2, initially projects this $(nAA, OH)$ matrix into the embedding space, resulting in a matrix of shape $(nAA, H)$, with $H$ representing the embedding size (in our case, $1280$).

Our decoding head resembles the NetSufP-3.0 architecture as described in Høie \textit{et al.} \cite{hoie2022netsurfp}. It features two separate CNN layers passed to a two-layer bidirectional Long Short-Term Memory (LSTM) network. The output is fed into a fully connected layer providing predictions for all residue and protein-level tasks. This design aims to extract information from the embeddings generated by the ESM-2 model, thus enhancing our model's capacity to utilise the representations offered by PLMs (Figure \ref{overview}). Consequently, post-embedding extraction, we apply a 1D-CNN to incorporate additional features to each residue, resulting in a matrix of shape $(nAA, H + O_{CNN} \times N_{CNN})$, where $O_{CNN}$ represents the number of CNN output channels (we assume uniformity across all CNN outputs, as is the case in PatchProt, where we use 2 CNNs with the output size of 32), and $N_{CNN}$ denotes the number of CNNs applied. To further enhance feature extraction, we apply bidirectional LSTM to these embeddings, resulting in a matrix of size $(nAA, 2 \times H_{LSTM})$, with $H_{LSTM}$ representing the hidden size of the LSTM (multiplied by $2$ due to the bidirectional nature). Concatenating the forward and backward representations (where $H_{LSTM} = 1024$ in PatchProt), we proceed to employ a linear layer for predicting the properties of each amino acid. Given that our multi-task model predicts multiple classes and values on both protein and residue levels simultaneously, the output is of shape $(nAA, nTasks)$. For all 13 prediction tasks, each has its dimensions. For instance, if classification among $8$ different classes is required, the output size would be $8$.

\subsection*{Combining global and local tasks}

Notably, our datasets contain additional global features. Hence, we developed a model which can predict both global features (LHP, species, and expression) and local features  ((L)HP, ASA, RSA, SS and disorder). To achieve this, for the global tasks, we simply sum the prediction of each amino acid for the specific task and use this as a prediction:

\begin{align}
Y = \displaystyle\sum_i^N y_i\label{eqS1}
\end{align}

The result is of shape $(1)$ since we predict one value per protein. This approach has one key advantage as it can emphasise the impact of each residue on the final label of the respective protein. This decomposition of the global features on a residue basis allows the interpretation of our results rather easily.

\subsection*{Scaling losses according to uncertainty}

When dealing with a multi-task learning scenario, where we aim to predict multiple properties using a single model, it is crucial to design a loss function that adequately accounts for the differences in tasks. These differences can arise from variations in scales or units, or even from the nature of tasks, such as classification or regression.

In the case of a regression model, we typically assume that the true values are normally distributed around their predicted counterparts \cite{kendall2018multitask}: $y_n \sim \mathcal{N}(f(x_n), \sigma^2)$. Where $\sigma^2$ represents the aleatoric uncertainty in our data (which is not reducible) and $f(x_n)$ denotes the model's prediction given the $n$th input from all inputs in $x$. We assume in our case that the uncertainty does not depend on the data (homoscedastic). During model optimisation, we would like to maximise the likelihood $p(\mathbf{y}|\mathbf{x})$. This likelihood can be written as: 

\begin{align}
\displaystyle\Pi_n^Np(y_n|f(x_n))\label{eqS2}
\end{align}

and respectively as: 

\begin{align}
\displaystyle\Pi_n^N \frac{1}{\sigma\sqrt{2\pi}}\exp^{\left(-\frac{1}{2}\left(\frac{y_n-f(x_n)}{\sigma}\right)^{2}\right)}\label{eqS3}
\end{align}

Where $N$ represents the total number of inputs. Typically, we take the log and multiply by $-1$ to minimise (easier to handle numerically). Additionally, we get rid of the constant, which brings us to the loss function \cite{kendall2018multitask, liebel2018auxiliary}: % (dividing by $N$ gives the MSE): 

\begin{align}
\frac{1}{2\sigma^2}L + \log(\sigma)\label{eqS4}
\end{align}

where:

\begin{align}
L = ||\mathbf{y},f(\mathbf{x})||^2 = \displaystyle\sum_n^N (y_n - f(x_n))^2\label{eqS5}
\end{align}

While in a single-task model, we remove the uncertainty term $\sigma$ (as we consider it a constant), in multi-task modelling we use the uncertainty to weigh different tasks. If we have two regression tasks with losses $L_1$ and $L_2$ similar to the one we computed above, but with different uncertainty $\sigma_1$ and $\sigma_2$ because they are not on the same scale/unit or simply because one of the tasks is noisier. We usually assume that the two tasks are independent: 

\begin{align}
p(\mathbf{y}_1,\mathbf{y}_2|f(\mathbf{x}),f(\mathbf{x}))  = p(\mathbf{y}_1|f(\mathbf{x}))p(\mathbf{y}_2|f(\mathbf{x}))\label{eqS6}
\end{align}

\begin{align}
L_{1,2} = \frac{1}{2\sigma_1^2}L_1 + \frac{1}{2\sigma_2^2}L_2 + \log(\sigma_1\sigma_2)\label{eqS7}
\end{align}

Where $\log(\sigma_1\sigma_2)$ is a regularisation term that prevents the uncertainty from increasing and masks one of the two tasks. This approach can also be used for classification tasks (where $\sigma$ represents the temperature, analogue of the uncertainty for categorical distribution). The multi-task loss function used in this paper is derived from \eqref{eqS7}:

\begin{align}
L_{multi}=\sum_{t\in \tau}\frac{L_t}{2\sigma_t^2}+\ln(1+\sigma_t^2),\label{eqS8}
\end{align}

where $\tau$ represents the set of tasks, $L_t$ is the loss function for task $t$ (mean squared loss and cross-entropy loss), and $\sigma_t$ is the uncertainty term for task $t$. To prevent negative loss values $\ln(1+\sigma_t^2)$ is administered in the approach instead of $log(\sigma_t)$ \cite{liebel2018auxiliary}. 

During optimisation, one way to implement this approach is to learn the uncertainty parameters $\sigma$ during training (as we cannot infer them before training) and to calculate the loss for each batch and adjust the uncertainty weights according to the optimisation objective. 

\subsection*{Fine-tuning strategy}

\begin{figure*}[htbp]
\centering
\includegraphics[width=17cm]{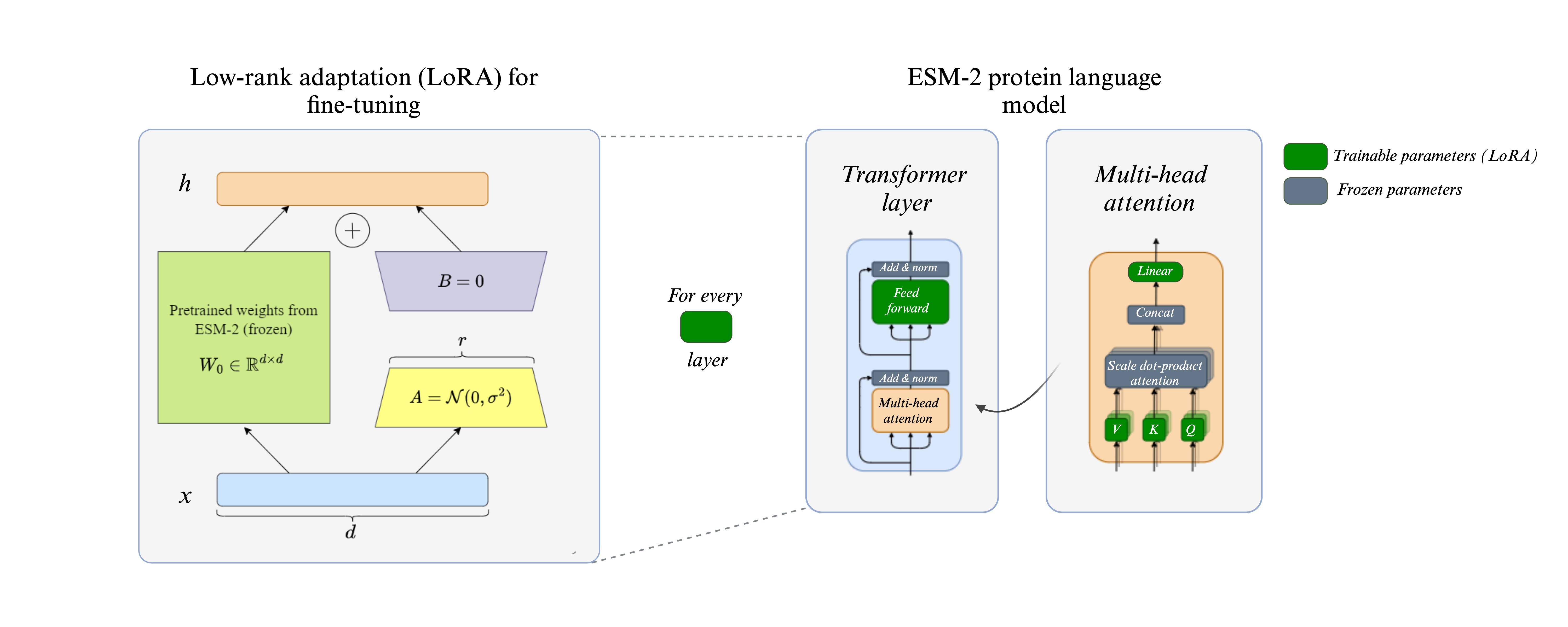}
\caption{\textbf{Fine-tuning strategy with Low-Rank Adaptation (LoRA).} 
To efficiently fine-tune the foundation model, we adopted recent advancements in parameter-efficient fine-tuning LoRa. In our approach, we applied LoRA to every linear layer within the original transformer architecture \cite{vaswani2017attention}, significantly reducing the number of updated parameters (to $2rd$ from the layer's original $d^2$). $W_0$ denotes the original weight matrix, decomposed into two lower-rank matrices, $A$ and $B$, with dimensions $r \times d$ and $d \times r$ respectively (see fine-tuning strategy section in Supplementary Information).}
\label{figsupp:LoRA}
\end{figure*}

To efficiently fine-tune the foundation model, we adopted recent advancements in parameter-efficient fine-tuning known as Low-Rank Adaptation (LoRA) \cite{hu2021LoRA} (Figure \ref{figsupp:LoRA}). With an expansion of LLMs, conventional methods of fine-tuning have grown impractical. LoRA has been demonstrated to significantly reduce the computational cost by freezing pre-trained model weights and introducing trainable rank decomposition matrices \cite{hu2021LoRA}.

The underlying principle of LoRA is based on the hypothesis that the change of weights during the fine-tuning process has an intrinsically low rank. This suggests that, rather than updating an entire weight matrix in each dense layer, only a few parameters are adjusted. Essentially, this hypothesis implies that most of the columns of the weight matrix are linearly dependent, eliminating the need to individually adjust each column.

In LoRA, $W_0$ denotes the original weight matrix of a specific layer within a pre-trained model, while $\Delta$W represents adjustments to weight changes to improve the layer's performance for new tasks. The final weights are obtained by adding $W_0$ and $\Delta$W matrices. The principal innovation of LoRA lies in decomposing the weight change matrix $\Delta$W into two lower-rank matrices, $A$ and $B$, with dimensions $r \times d$ and $d \times r$ respectively:

\begin{align}
A\in \mathbb{R}^{r\times d},\: 
B\in \mathbb{R}^{d\times r},\: r\ll d,
\label{eqs9}
\end{align}

This approach significantly reduces the number of updated parameters (to $2rd$ from the layer's original $d^2$), thereby enhancing the efficiency of the fine-tuning process. These updates are applied through residual connections, allowing us to modify the model's behaviour with minimal changes to its pre-trained weights. The adapted output $h$ for a new input $x$ is computed as follows:

\begin{align}
h=W_0x + \Delta Wx=W_0x + BAx,
\label{eqs10}
\end{align}

Here, $\Delta W = BA$ represents the weight adjustments through the low-rank decomposition, where only matrices $A$ and $B$ are updated, improving the efficiency of the training while maintaining the integrity of the pre-trained model.

In our approach, we applied LoRA to every linear layer within the original transformer architecture \cite{vaswani2017attention}, targeting not only queries, keys and values matrices but also the projection layer in the multi-head attention and the feed-forward network in the transformer as shown in Figure \ref{figsupp:LoRA}.

\subsection*{Handling long sequences}

It is computationally expensive to generate embeddings for long FASTA sequences. Here we propose an approach to parse long sequences, by introducing a new parameter to achieve a better representation of long inputs. As in NetSurfP-3.0 \cite{hoie2022netsurfp}, we divide the input into several parts of equal lengths of 1,048 amino acids. Afterwards, instead of truncating only the end of the previous part, we truncate both embeddings and assemble the results. 

\subsection*{Batch size correction}

Combining global and local tasks is challenging in terms of batch sizes. Since every amino acid residue is a training sample for local features the model requires way fewer sequences to be trained compared to the global features, when it has to predict a single value per protein. Furthermore, as computing and storing the embeddings of large proteins is computationally heavy, a single GPU can only be used for a smaller batch size. To ensure that we learn the global features adequately, we need to maximise the batch size. For this, we applied gradient optimisation techniques to allow a greater batch size on a single GPU, namely, gradient accumulation and gradient checkpointing. Using these techniques we can increase our batch size of $2$ (2 proteins) to $18$. Using gradient checkpointing on half of the transformers allows us to increase the batch size to $3$ and gradient accumulation can be used to increase the batch size (we accumulate the gradient over $6$ batches) resulting in a virtual batch size of $6 \times 3 = 18$ molecules per batch.

\section*{Supplementary tables}

\begin{table*}[htbp]
    \centering
    \small
    \begin{tabular}{ccccccccc}
        \hline
       \multirow{2}{*}{Dataset} & \multirow{3}{*}{Task} & \multirow{3}{*}{Feature} & \multirow{3}{*}{Source} & Train & Test & Test & Test & Overlap \\
        &  &  &  &  HHBlits &  CASP12 &  CB513 &  TS115  &  with the original \\
       Size &  &  &  & 10,848 & 21 & 513 & 115 & train \& test datasets \\
       \hline
       \multirow{2}{*}{Original} & \multirow{2}{*}{Local} & Q8, Q3, RSA, ASA, & \multirow{2}{*}{DSSP} & \multirow{2}{*}{10,848} & \multirow{2}{*}{21} & \multirow{2}{*}{513} & \multirow{2}{*}{115} & \multirow{2}{*}{-} \\ 
       && $\phi$, $\psi$, Disorder &  & &  &\\ 
       Added & Global & TASA, THSA & DSSP & 10,848  & 21 & 513 & 115  & 100\%\\ 
       Added & Both & (L)HP & MolPatch & 9,991 & 20 & 470 & 113 & 100\%\\
       Added & Global & NX  & HPA & 579 & 1 & 31 & 7 & 100\%\\
       Added & Global & SP  & RCSB & 3,528 & 4 & 181 & 115& 100\%\\
       \hline
    \end{tabular}
    \caption{Training and test data with additional features. Q8 - Eight-state secondary structure, Q3 - three-state secondary structure, RSA - relative solvent-accessible area, ASA - absolute solvent-accessible area, TASA - total accessible surface area, THSA - total hydrophobic surface area, LHP - largest hydrophobic patch, NX - normalised expression, SP - species.}
    \label{tabsupp:datasets}
\end{table*}

\begin{table*}[htbp]
\small
\begin{tabular}{@{\extracolsep\fill}lllllllllll@{\extracolsep\fill}}
\hline
Test & \multirow{2}{*}{LoRA} & \multirow{2}{*}{Model} & RSA $\uparrow$  & ASA $\uparrow$  & Q8  $\uparrow$ & Q3  $\uparrow$  & Dis $\uparrow$ & Dis $\downarrow$ & Phi $\downarrow$ & Psi $\downarrow$ \\
dataset &  & & (PCC) & (PCC) & (ACC) & (ACC) & (MCC) & (FNR)  & (MAE) & (MAE)\\
\hline

\multirow{8}{*}{CASP12}
&  & NetSurfP-2.0 & 0.728 & 0.739 & 0.699 & 0.810 & 0.653 & 0.015 & 20.90 & 32.80\\
&  & NetSurfP-3.0 & 0.707 & 0.722 & 0.669 & 0.791 & 0.621 & 0.024 & 21.25 & 33.92 \\
& $\times$ & ESM-2 & 0.710 & 0.717 & 0.653 & 0.785 & 0.543 & \textbf{0.013} & 21.49 & 33.48 \\
& $\surd$ & ESM-2 & 0.707 & 0.724 & 0.666 & 0.777 & 0.559 & 0.021 & 20.60 & 32.50 \\
& $\times$ & PatchProt (SSE) & \textbf{0.740} & \textbf{0.748} & \textbf{0.695} & \textbf{0.817} & \textbf{0.658} & 0.026 & \textbf{20.20} & \textbf{30.95} \\
& $\surd$ & PatchProt (SSE) & 0.720 & 0.735 & 0.683 & 0.792 & 0.579 & 0.029 & 20.39 & 31.78 \\
& $\times$ & PatchProt (all tasks) & 0.730 & 0.738 & 0.667 & 0.799 & 0.583 & 0.024 & 20.80 & 31.33 \\
& $\surd$ & PatchProt (all tasks) & 0.724 & 0.741 & 0.685 & 0.795 & 0.592 & 0.032 & 20.42 & 32.42 \\
\hline

\multirow{8}{*}{CB513}
&  & NetSurfP-2.0 & 0.791 & 0.804 & 0.713 & 0.845 & - & - & 20.35 & 29.04\\
&  & NetSurfP-3.0 & 0.793 & 0.810 & 0.711 & 0.846 & - & - & 20.22 & 29.25\\
& $\times$ & ESM-2 & 0.791 & 0.804 & 0.682 & 0.836 & - & - & 20.76 & 30.28 \\
& $\surd$ & ESM-2 & 0.803 & 0.817 & 0.724 & 0.859 & - & - & 19.34 & 26.60 \\
& $\times$  & PatchProt (SSE) & 0.811 & 0.823 & 0.724 & 0.860 & - & - & 19.47 & 26.73 \\
& $\surd$ & PatchProt (SSE) & \textbf{0.816} & \textbf{0.828} & 0.737 & \textbf{0.868} & - & - & 18.93 & \textbf{25.56} \\
& $\times$ & PatchProt (all tasks)  & 0.809 & 0.821 & 0.704 & 0.855 & - & - & 19.93 & 27.87 \\
& $\surd$ & PatchProt (all tasks) & \textbf{0.816} & \textbf{0.828} & \textbf{0.738} & \textbf{0.868} & - & - & \textbf{18.83} & 25.72 \\
\hline

\multirow{8}{*}{TS115}
&  & NetSurfP-2.0 & 0.771 & 0.793 & 0.740 & 0.849 & 0.624 & \textbf{0.013} & 17.40 & 26.80 \\
&  & NetSurfP-3.0 & 0.776 & 0.799 & 0.749 & 0.856 & 0.662 & 0.015 & 17.16 & 25.80 \\
& $\times$ & ESM-2 & 0.772 & 0.790 & 0.719 & 0.844 & 0.605 & \textbf{0.013} & 17.76 & 27.00 \\
& $\surd$ & ESM-2 & 0.785 & 0.805 & 0.753 & 0.858 & 0.646 & 0.016 & 16.76 & 24.42 \\
& $\times$ & PatchProt (SSE) & 0.796 & 0.812 & 0.757 & 0.867  & \textbf{0.667} & 0.016 & 16.67 & 23.75 \\
& $\surd$ & PatchProt (SSE) & 0.794 & 0.813 & 0.763 & 0.869 & 0.650 & 0.014 & 16.37 & \textbf{23.59} \\
& $\times$ & PatchProt (all tasks) & 0.792 & 0.809 & 0.739 & 0.861 & 0.650 & 0.014 & 17.03 & 24.83 \\
& $\surd$ & PatchProt (all tasks) & \textbf{0.799} & \textbf{0.817} & \textbf{0.765} & \textbf{0.871} & 0.649 & 0.016 & \textbf{16.24} & 23.67 \\
\hline

\end{tabular}
\caption{\textbf{Model performance when applying ESM-2 embeddings to predict protein local structure.} Comparison of NetSurfP-2.0, NetSurfP-3.0, and our model - PatchProt on the CB513, TS115 and CASP12 datasets. Performance values for the NetsurfP models are reported as stated in the latest publication \cite{hoie2022netsurfp}. ESM-2 consists of the ESM-2 embedding model with linear layers at its end for making predictions. SSE - secondary structure element (the model was only trained on basic secondary structure component tasks). SSE + auxiliary tasks additionally include global tasks (TASA, THSA), (largest) hydrophobic patches (both global and local), species, and expression. Each column reports an output variable with the same corresponding metrics reported in the previous study \cite{hoie2022netsurfp} for benchmarking purposes: Pearson correlation coefficient (PCC), accuracy (ACC), Matthews correlation coefficient (MCC), false negative rate (FNR) and mean absolute error (MAE). Up- and down-facing arrows indicate metrics for which an improvement represents larger or lower values. For each dataset and prediction task, the values corresponding to the best performance are shown in bold.
\label{tabsupp:basic_tasks_SI}}%
\end{table*}

\newpage
\section*{Supplementary figures}

\begin{figure*}[htbp]
\centering
\includegraphics[width=17cm]{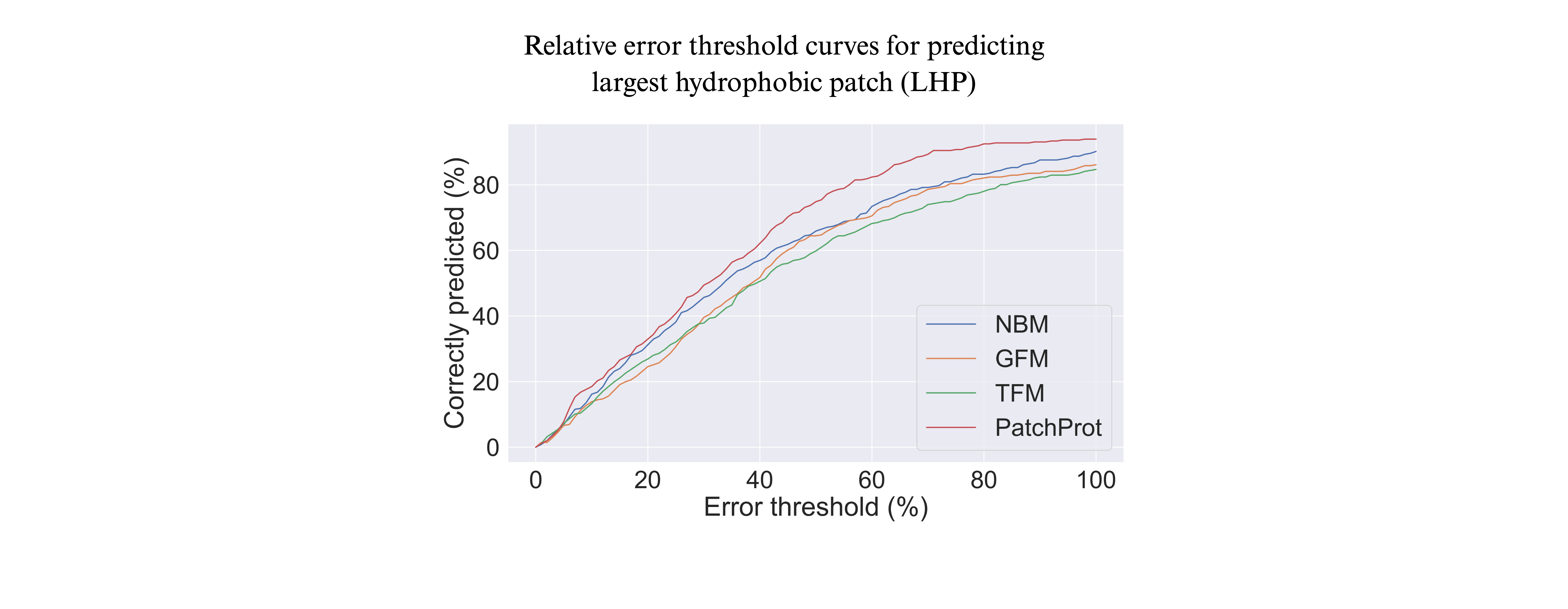}
\caption{\textbf{Benchmarking global largest hydrophobic predictions (LHP).} The Accuracy of the predictions of the largest patch hydrophobic surface area was compared using threshold curves. Global predictions by PatchProt are benchmarked against other methods, including the three-feature model (TFM), which uses the sequence length, number of hydrophobic amino acids and number of hydrophilic amino acids as input features \cite{kuhn2008building}. The global feature model (GFM) trained on 31 global features using an XGBoost regressor \cite{chen2016xgboost}. NetSurfP-2.0-based model (NBM), which is a random forest model trained using the relative and total hydrophobic surface area values (THSA, RHSA) predicted by NetSurfP-2.0, since the LHP cannot be calculated from NetSurfP-2.0 predictions directly \cite{van2022sticky}. The fraction of correctly predicted proteins within a certain error margin for each of the methods is shown as calculated over the test set. The test set and the threshold curve calculations were replicated from the previous study \cite{van2022sticky}. Importantly, the large fraction of proteins in the test set were used to train PatchProt. For a fair comparison, the overlapping proteins were removed and the curves were calculated for the rest of the proteins in the test set (n=346).}
\label{figsupp:threshold_curve}
\end{figure*}

\end{document}